\documentclass[aps,prd,notitlepage,amssymb,amsmath,floatfix,nofootinbib,superscriptaddress]{revtex4-1}

\usepackage{epsfig}
\usepackage{bm}
\usepackage{amssymb}
\usepackage{amsmath}
\usepackage{color}
\usepackage{slashed}
\usepackage{subfigure}
\usepackage[colorlinks,linkcolor=blue,anchorcolor=black,citecolor=blue]{hyperref}
\allowdisplaybreaks[4]
\usepackage{comment}

\begin{document}

\title{Dihadron helicity correlation in photon-nucleus collisions}

\author{Zhao-Xuan Chen}
\affiliation{Institute of Frontier and Interdisciplinary Science, Key Laboratory of Particle Physics and Particle Irradiation (MOE), Shandong University, Qingdao, Shandong 266237, China}

\author{Hui Dong}
\email{hdong@sdu.edu.cn}
\affiliation{Institute of Frontier and Interdisciplinary Science, Key Laboratory of Particle Physics and Particle Irradiation (MOE), Shandong University, Qingdao, Shandong 266237, China}

\author{Shu-Yi Wei}   
\email{shuyi@sdu.edu.cn}
\affiliation{Institute of Frontier and Interdisciplinary Science, Key Laboratory of Particle Physics and Particle Irradiation (MOE), Shandong University, Qingdao, Shandong 266237, China}

\begin{abstract}
The helicity correlation of two back-to-back hadrons is a powerful tool that makes it possible to probe the longitudinal spin transfer, $G_{1L}$, in unpolarized hadronic collisions. In this work, we investigate the helicity correlation of back-to-back dihadrons produced in photon-nucleus collisions with both spacelike and quasireal photons and explore its potential in understanding the flavor dependence of spin-dependent fragmentation functions. We present helicity amplitudes of partonic scatterings with both virtual and real photons and make numerical predictions for the dihadron helicity correlations at the future Electron Ion Collider experiment and the current RHIC/LHC ultraperipheral collision experiment. Future experimental measurements can also illuminate the fragmentation function of circularly polarized gluons. 
\end{abstract}

\maketitle

\section{Introduction}

The longitudinal spin transfer of fragmentation functions, $G_{1L}$, encodes information on how the polarization of the fragmenting parton is inherited by final-state hadrons \cite{Metz:2016swz, Chen:2023kqw, Boussarie:2023izj} during the hadronization. Therefore, it is an important quantity in understanding the hadronization mechanism of high-energy partons, providing a new dimension beyond the momentum distribution \cite{Binnewies:1994ju, Kniehl:2000fe, Kneesch:2007ey, Kretzer:2000yf, Albino:2005me, Albino:2005mv, Albino:2008fy, deFlorian:2007aj, deFlorian:2007ekg, Hirai:2007cx, Aidala:2010bn, deFlorian:2014xna, deFlorian:2017lwf, Bertone:2017tyb, Khalek:2021gxf, arXiv:1905.03788, arXiv:2101.04664, arXiv:2210.06078}. Since the longitudinal spin transfer describes the fragmentation of polarized partons, we are not surprised by the fact that it was first investigated by the LEP experiment \cite{Augustin:1978wf,Gustafson:1992iq,ALEPH:1996oew, OPAL:1997oem} and the polarized semi-inclusive deeply inelastic experiments \cite{E665:1999fso,HERMES:1999buc,NOMAD:2000wdf, NOMAD:2001iup,HERMES:2006lro,COMPASS:2009nhs}. However, the quantitative study of this polarized fragmentation function is far from being satisfactory. Currently, only one parametrization \cite{deFlorian:1997zj} for $G_{1L}$ of $\Lambda$ is available in the market, which stopped updating after 2000 since there are no new experimental data available. The flavor dependence of $G_{1L}$ remains unsolved.

The dihadron longitudinal polarization correlation (refereed to as the {\it helicity correlation} in the rest of this paper) \cite{Chen:1994ar} has been proposed as a novel observable that allows us to probe the longitudinal spin transfer even in the Belle experiment. However, no such measurement has ever been conducted yet in $e^+e^-$ annihilation experiments. Equipped with the helicity amplitude method, recent works \cite{Zhang:2023ugf, Li:2023qgj} have also extended this approach to unpolarized hadronic collisions. This progress allows us to avoid the contamination of polarized parton distribution functions, which are also poorly constrained so far, in the study of $G_{1L}$ in hadronic collisions. Moreover, albeit in a different context, the polarization correlation of dihadrons has indeed been measured at RHIC experiment \cite{Gong:2021bcp,Vanek:2023oeo,Tu:2023few} demonstrating that such a goal is indeed achievable. In light of the huge amount of experimental data that have already collected by RHIC, Tevatron and the LHC experiments, it is promising to elevate the quantitative study of the polarized fragmentation function $G_{1L}$ in the imminent future. 

In the helicity amplitude approach, one can factorize the partonic hard scattering from the nonperturbative physics in the helicity basis. In general, we obtain multiple helicity amplitudes since there are quite a few helicity combinations of initial and final state partons. However, some of the combinations do not contribute and the rest can usually be connected with each other. Therefore, the final formula is not as complicated as one initially imagined. Furthermore, since we are interested in the longitudinal polarization, it is straightforward to employ the helicity amplitude approach. Thanks to the number density interpretation of leading twist fragmentation functions, one can simply convolute the helicity-dependent partonic cross-sections and the helicity-dependent fragmentation functions to derive the hadronic cross-section that can be used to compute the dihadron helicity correlation. More applications of the helicity amplitude approach can be found in \cite{DeCausmaecker:1981wzb,Gastmans:1990xh,Anselmino:2005sh, Anselmino:2011ch, DAlesio:2021dcx}.

With the next-generation electron-ion collider (EIC) \cite{AbdulKhalek:2021gbh} on the horizon, it is timely to investigate the opportunity of exploring the longitudinal spin transfer in photon-nucleus collisions. In contrast with the previous proposals using polarized beams \cite{Jaffe:1996wp, Kotzinian:1997vd, deFlorian:1997kt, Ashery:1999am, deFlorian:1998ba, Jager:2002xm, Xu:2002hz, Xu:2005ru}, we focus on unpolarized collisions. Because of the parity conservation, it is not possible to study the longitudinal spin transfer in the single-inclusive hadron production process. However, it becomes feasible in the back-to-back dijet production process. From each jet, we select a $\Lambda$ or $\bar\Lambda$ hyperon and investigate the helicity correlation of these two hyperons residing in different jets. Effectively, we consider the following process: $\gamma^* + A \to {\rm jet}_1 (\to \Lambda/\bar \Lambda) + {\rm jet}_2 (\to \Lambda/\bar \Lambda) +X$. The study is akin to those in Refs. \cite{Zhang:2023ugf, Li:2023qgj}. While neither of the final state partons is individually polarized, their helicities are correlated. This partonic level helicity correlation is eventually translated into the polarization correlation of final state hadrons. The benefit of including photon-nucleus process in the global analysis is manifest. The flavor dependence of fragmentation functions can hardly be determined by a single experiment, if not impossible. The photon-nucleus collision offers a new candidate that can help to constrain free parameters in fragmentation functions.

On the other hand, the ultraperipheral relativistic heavy-ion collision (UPC) also offers a clean platform of studying polarized fragmentation functions. This is different from the central nucleus-nucleus collisions where the partonic hard scatterings are tainted by the jet-medium interactions. The quark gluon plasma does not form in UPC. There are two sorts of UPC processes. The first one is that both nuclei remain unbroken, where the partonic interactions are $\gamma + \gamma$ and $\gamma$+Pomeron scatterings. This process has already been studied in Ref.~\cite{Li:2023qgj}. Another process is that only one of the nuclei remains unbroken, where the underlying events are $\gamma + A \to {\rm jet}_1 (\to \Lambda/\bar \Lambda) + {\rm jet}_2 (\to \Lambda/\bar \Lambda) +X$. This process has not been studied yet and it resembles the dijet production process in the EIC experiment. The difference lies in the virtuality of the projectile photon. While the photons in the EIC experiment are spacelike with a large virtuality, those in UPC are quasireal with typical virtuality at around 30 MeV. Therefore, we can safely utilize the real photon approximation in the theoretical calculation. Nonetheless, a systemic investigation in the current UPC experiments still offers a complementary role in understanding the $G_{1L}$ fragmentation function.

The rest of the paper is organized as follows. In Sec. II, we lay out the formulas in the theoretical calculation. Particularly, we present the helicity-dependent partonic cross-sections with a virtual photon. Convoluting with the nonperturbative inputs, we obtain the hadronic cross-sections for EIC and UPC experiments. In Sec. III, we show our numerical predictions for the dihadron helicity correlations employing the DSV parameterization. A summary is given in Sec. IV.

\section{Formulas}

In this section, we first lay out the helicity amplitudes for $\gamma^{*} + g \to q + \bar q$ and $\gamma^{*} + q \to g + q$ with a virtual photon. Taking the photon virtuality $Q = 0$ limit, we can immediately obtain the helicity amplitudes for the UPC process. Then, we present the formulas to compute dihadron helicity correlations at EIC and UPC experiments. 

\subsection{Kinematics}

The leading order partonic hard scattering is denoted as $\gamma^* (q) + b (p_2) \to c (p_3, \lambda_c) + d (p_4, \lambda_d)$ with $q$ and $p_i$ denoting the four-momenta of the virtual photon and the corresponding parton and $\lambda_{c,d}$ denoting the helicity. We can define Mandelstam variables as
\begin{align}
&
\hat s = (q + p_2)^2 = 2 q \cdot p_2 - Q^2,
\\
& 
\hat t = (q - p_3)^2 = - 2 q \cdot p_3 - Q^2,
\\
&
\hat u = (q - p_4)^2 = - 2 q \cdot p_4 - Q^2,
\end{align}
where we have employed $q^2 = - Q^2$ to arrive at final expressions. Because of the momentum conservation, it is straightforward to verify that $\hat s + \hat t + \hat u + Q^2 = 0$.

Since the projectile photon is spacelike, we have to consider both longitudinal and transverse polarizations of the virtual photon. Therefore, we arrive at transverse and longitudinal partonic cross-sections given by
\begin{align}
&
\sigma_T = \frac{1}{2} \sum_{\lambda_{\gamma} =\pm 1} \epsilon_{\mu} (\lambda_{\gamma}) \epsilon_{\nu}^* (\lambda_{\gamma}) \sigma^{\mu\nu}, \label{eq:tv}
\\
&
\sigma_L =  \epsilon_{\mu} (\lambda_{\gamma}=0) \epsilon_{\nu}^* (\lambda_{\gamma}=0) \sigma^{\mu\nu}, \label{eq:lv}
\end{align}
Here $\lambda_{\gamma}$ is the photon helicity and $\epsilon_\mu$ is the polarization vector of the virtual photon. $\sigma^{\mu\nu}$ is the hadronic tensor with $\mu$ and $\nu$ indices coming from the quark-photon interaction vertices. The transverse and longitudinal polarization vectors are given by 
\begin{align}
&
\epsilon_\mu (\lambda_\gamma = \pm 1) = \frac{1}{\sqrt{2}} (0, 0, 1, \pm i),
\\
&
\epsilon_\mu (\lambda_\gamma = 0) = \frac{1}{Q} (q^+,  - q^-, \bm{0}_\perp),
\end{align}
in the light-cone coordinates. Here we have already utilized the fact that the spacelike photon moves along the $z$ axis. However, the above expressions for polarization vectors are difficult to implement in the computer language. Therefore, we usually parametrize them as \cite{Gastmans:1990xh, Jezuita-Dabrowska:2002tjg}
\begin{align}
& \epsilon_\mu (\lambda_\gamma = \pm 1) = 
\frac{\frac{(p_3 \cdot q) (p_2\cdot q) + Q^2 p_2 \cdot p_3}{p_2\cdot q} p_{2\mu} + p_2 \cdot p_3 q_\mu - (p_2 \cdot q) p_{3\mu} \pm i \varepsilon_{\mu\alpha \beta \gamma} p_3^\alpha p_2^\beta q^\gamma}{\sqrt{4(p_2 \cdot p_3) (p_2 \cdot q) (p_3\cdot q) + 2Q^2 p_2 \cdot p_3}},
\\
& \epsilon_\mu (\lambda_\gamma = 0) = \frac{Q}{p_2 \cdot q} \left( p_{2\mu} + \frac{p_2\cdot q}{Q^2} q_\mu \right).
\end{align}
Here, $p_2$ moves along the $-\bm{e}_z$ direction and $p_3$ contains a transverse component. This parametrization differs with Eqs.~(\ref{eq:tv}-\ref{eq:lv}) by a gauge transform and it also satisfies
\begin{align}
& q \cdot \epsilon_{L, T} = 0,
&&
\epsilon_T (\lambda_\gamma) \cdot \epsilon_{T}^*  (\lambda_\gamma) = -1,
&&
\epsilon_L \cdot \epsilon_{L}^* = +1.
\end{align}
Since the final cross section is gauge invariant, we have the liberty to utilize either expression. 

Moreover, employing the completeness relation $-g_{\mu\nu} - \frac{q_\mu q_\nu}{Q^2}  = \sum_{\lambda_\gamma = \pm 1} \epsilon_\mu (\lambda_\gamma) \epsilon_\nu^* (\lambda_\gamma) - \epsilon_\mu (\lambda_\gamma = 0) \epsilon_{\nu}^* (\lambda_\gamma = 0)$ and the current conservation $q_\mu \sigma^{\mu\nu} = q_\nu \sigma^{\mu\nu} = 0$, it is convenient to define $\sigma_{\Sigma}$ as
\begin{align}
\sigma_{\Sigma} = - g_{\mu\nu} \sigma^{\mu\nu}  = 2 \sigma_T - \sigma_L. \label{eq:cs-sum}
\end{align}
The subscript $\Sigma$ merely indicates that $\sigma^{\mu\nu}$ is contracted by a $-g_{\mu\nu}$. $\sigma_{\Sigma}$ defined above will only be used to compare our unpolarized results with those in traditional textbooks, such as Ref.~\cite{Field:1989uq}.

Last but not least, in this work, we have only considered the diagonal contributions, i.e., the photon helicities in the amplitude and the conjugate amplitude are the same. Once mixtures between different photon helicities are taken into account, the interference terms give birth to $\cos \phi$ and $\cos2\phi$ dependent contributions with $\phi$ being the azimuthal angle between leptonic plane and the hadronic plane \cite{Jezuita-Dabrowska:2002tjg}. These azimuthal angle dependent terms also introduce a host of interesting topics that are beyond the scope of this work. Since we will eventually integrate over the azimuthal angle, these interference terms simply vanish and thus are not considered.

\subsection{Helicity-dependent partonic cross sections}


\begin{figure}[htb]
\includegraphics[width=0.25\textwidth]{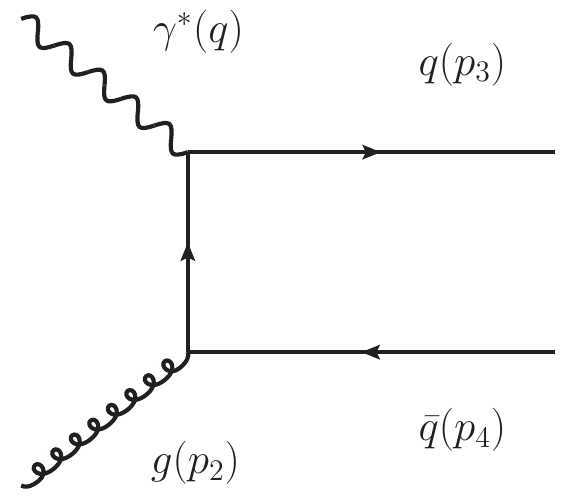}
\phantom{XXXX}
\includegraphics[width=0.25\textwidth]{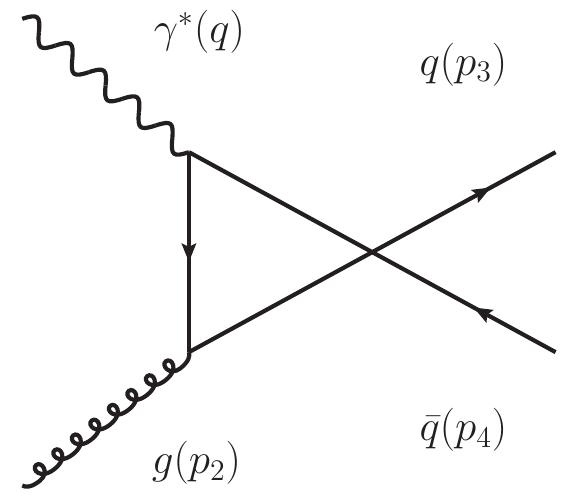}
\caption{Feynman diagrams of the $\gamma^* + g \to q + \bar q$ channel.}
\label{fig:fey-gg}
\end{figure}

We start with the $\gamma^* (q) + g (p_2) \to q (p_3, \lambda_q) + \bar q (p_4, \lambda_{\bar q})$ channel which is shown in Fig.~\ref{fig:fey-gg}. Since we only study the final state helicity correlation in the unpolarized collisions, we average over the spin degrees of freedom of initial state partons. The final-state helicity-dependent partonic cross-sections of this channel thus read
\begin{align}
&
\frac{d\hat\sigma^{\gamma_T^* g \to q\bar q}_{+-}}{d\hat t} = \frac{d\hat\sigma^{\gamma_T^* g \to q\bar q}_{-+}}{d\hat t}
= \frac{1}{2} \frac{\pi \alpha \alpha_s e_q^2}{(\hat s + Q^2)^2} 
\left[
\frac{\hat u}{\hat t} + \frac{\hat t}{\hat u} 
- \frac{2Q^2 \hat s}{\hat u \hat t} + \frac{4Q^2 \hat s}{(\hat s + Q^2)^2}
\right],
\\
&
\frac{d\hat\sigma^{\gamma_L^* g \to q\bar q}_{+-}}{d\hat t} = \frac{d\hat\sigma^{\gamma_L^* g \to q\bar q}_{-+}}{d\hat t}
= \frac{\pi \alpha \alpha_s e_q^2}{(\hat s + Q^2)^2} 
\frac{4 Q^2 \hat s}{ (\hat s + Q^2)^2}.
\end{align}
Here, $+-$ or $-+$ in the subscript denotes the helicity combination of $q$ and $\bar q$. The other combinations ($++$ and $--$) simply vanish due to the helicity conservation. 

The unpolarized partonic cross-sections can be obtained by summing over all possible helicity combinations. We find
\begin{align}
&
\frac{d\hat\sigma^{\gamma_T^* g \to q\bar q}_{\rm unpolarized}}{d\hat t} = 
\frac{\pi \alpha \alpha_s e_q^2}{(\hat s + Q^2)^2} 
\left[
\frac{\hat u}{\hat t} + \frac{\hat t}{\hat u} 
- \frac{2Q^2 \hat s}{\hat u \hat t} + \frac{4Q^2 \hat s}{(\hat s + Q^2)^2}
\right],
\\
&
\frac{d\hat\sigma^{\gamma_L^* g \to q\bar q}_{\rm unpolarized}}{d\hat t} = 
\frac{\pi \alpha \alpha_s e_q^2}{(\hat s + Q^2)^2} 
\frac{8 Q^2 \hat s}{ (\hat s + Q^2)^2},
\end{align}
and $\hat\sigma_\Sigma$ defined according to Eq.~(\ref{eq:cs-sum}) agrees with that in Ref.~\cite{Field:1989uq}.

Furthermore, taking the $Q^2 \to 0$ limit, we obtain the partonic cross-sections for the real photon which are given by
\begin{align}
&
\frac{d\hat\sigma^{\gamma_T g \to q\bar q}_{+-}}{d\hat t} = \frac{d\hat\sigma^{\gamma_T g \to q\bar q}_{-+}}{d\hat t}
= \frac{\pi \alpha \alpha_s e_q^2}{2\hat s^2} 
\left[
\frac{\hat u}{\hat t} + \frac{\hat t}{\hat u} 
\right],
\label{eq:realphotong}
\\
&
\frac{d\hat\sigma^{\gamma_L g \to q\bar q}_{+-}}{d\hat t} = \frac{d\hat\sigma^{\gamma_L g \to q\bar q}_{-+}}{d\hat t}
= 0.
\end{align}
As expected, the cross-section of the unphysical longitudinally polarized photon disappears, while that of the transversely polarized photon given in Eq.~(\ref{eq:realphotong}) exactly reproduces the result presented in Ref.~\cite{Gastmans:1990xh}.

\begin{figure}[htb]
\includegraphics[width=0.25\textwidth]{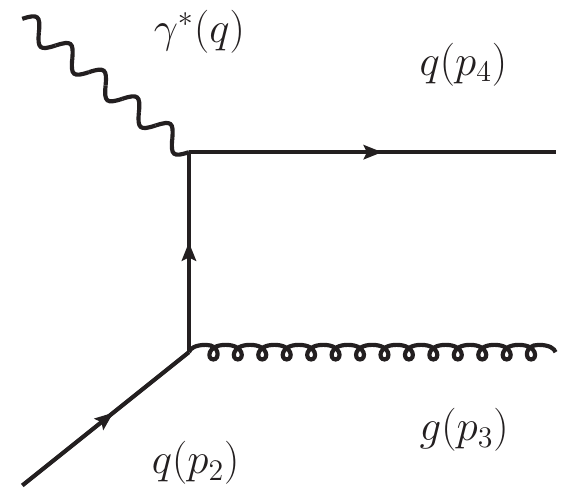}
\phantom{XXXX}
\includegraphics[width=0.25\textwidth]{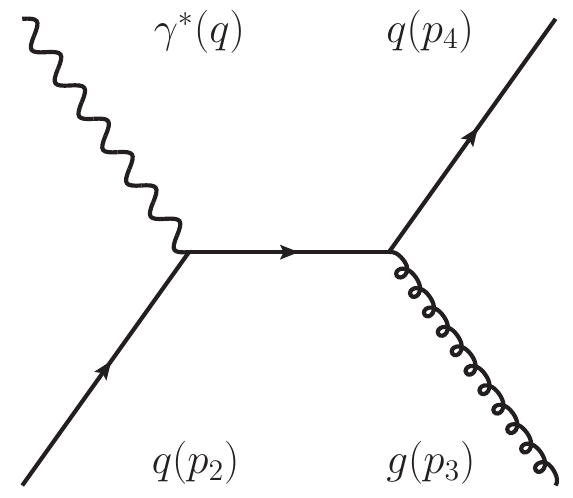}
\caption{Feynman diagrams of the $\gamma^* + q \to g + q$ channel.}
\label{fig:fey-gq}
\end{figure}

The second channel is the $\gamma^*(q) + q (p_2) \to g (p_3, \lambda_g) + q (p_4, \lambda_q)$ diagram illustrated in Fig.~\ref{fig:fey-gq}. For the virtual photon, the helicity dependent cross-sections are given by
\begin{align}
&
\frac{d\hat\sigma^{\gamma_T^* q \to g q}_{++}}{d\hat t} = \frac{d\hat\sigma^{\gamma_T^* q \to g q}_{--}}{d\hat t}
= 
- \frac{4}{3} 
\frac{\pi \alpha \alpha_s e_q^2}{(\hat s + Q^2)^2} \frac{(\hat s + Q^2)^2}{\hat s \hat u},
\label{eq:cs-1}
\\
&
\frac{d\hat\sigma^{\gamma_T^* q \to g q}_{+-}}{d\hat t} = \frac{d\hat\sigma^{\gamma_T^* q \to g q}_{-+}}{d\hat t}
= 
- \frac{4}{3} \frac{\pi \alpha \alpha_s e_q^2}{(\hat s + Q^2)^2}
\left[ \frac{(\hat u + Q^2)^2}{\hat s \hat u}  + \frac{2Q^2 \hat t}{(\hat s + Q^2)^2} \right],
\\
&
\frac{d\hat\sigma^{\gamma_L^* q \to g q}_{++}}{d\hat t} = \frac{d\hat\sigma^{\gamma_L^* q \to g q}_{--}}{d\hat t}
= 0,
\\
&
\frac{d\hat\sigma^{\gamma_L^* q \to g q}_{+-}}{d\hat t} = \frac{d\hat\sigma^{\gamma_L^* q \to g q}_{-+}}{d\hat t}
= 
- \frac{4}{3} 
\frac{\pi \alpha \alpha_s e_q^2}{(\hat s + Q^2)^2}
\frac{4Q^2 \hat t}{(\hat s + Q^2)^2}.
\label{eq:cs-4}
\end{align}
It is intriguing to note that the helicities of final state quark and gluon also take the opposite signs for the longitudinal photon. The cross-sections with same sign helicities vanish. On the other hand, for the transverse photon, both same sign and opposite sign combinations contribute. A similar feature also emerges in the deeply virtual Compton scattering $\gamma_{L/T}^* + q \to \gamma + q$. Ref.~\cite{Huang:2003uy} shows that the amplitude of $\gamma^*_L + q \to \gamma (\lambda_\gamma = +) + q (\lambda_q = +)$ also vanishes. Furthermore, the three-gluon vertex does not appear at the leading order. Therefore, our helicity amplitudes in Eqs.~(\ref{eq:cs-1}-\ref{eq:cs-4}) for the $\gamma^* + q \to g + q$ channel completely agree with their results \cite{Huang:2003uy} for the $\gamma^* + q \to \gamma + q$ channel.

Summing over the final state spin degree of freedom, we obtain the unpolarized partonic cross-sections given by
\begin{align}
&
\frac{d\hat\sigma^{\gamma_T^* q \to g q}_{\rm unpolarized}}{d\hat t}
= - \frac{8}{3} 
\frac{\pi \alpha \alpha_s e_q^2}{(\hat s + Q^2)^2} \left[ \frac{\hat s}{\hat u} + \frac{\hat u}{\hat s} - \frac{2Q^2 \hat t}{\hat s \hat u} + \frac{2Q^2 \hat t}{(\hat s + Q^2)^2} \right],
\\
&
\frac{d\hat\sigma^{\gamma_L^* q \to g q}_{\rm unpolarized}}{d\hat t}
= 
- 
\frac{8}{3} 
\frac{\pi \alpha \alpha_s e_q^2}{(\hat s + Q^2)^2}
\frac{4Q^2 \hat t}{(\hat s + Q^2)^2}
.
\end{align}
One can easily verify that $\hat\sigma_\Sigma$ of this channel reproduces that in Ref.~\cite{Field:1989uq}.

Again, taking the $Q^2 \to 0$ limit, the longitudinal cross-section vanishes and the transverse cross-section coincides with those in Ref.~\cite{Gastmans:1990xh}. We obtain
\begin{align}
&
\frac{d\hat\sigma^{\gamma_T q \to g q}_{++}}{d\hat t} = \frac{d\hat\sigma^{\gamma_T q \to g q}_{--}}{d\hat t}
= 
- \frac{4}{3} \frac{\pi \alpha \alpha_s e_q^2}{\hat s^2} \frac{\hat s}{\hat u} ,
\\
&
\frac{d\hat\sigma^{\gamma_T q \to g q}_{+-}}{d\hat t} = \frac{d\hat\sigma^{\gamma_T q \to g q}_{-+}}{d\hat t}
= 
- \frac{4}{3} \frac{\pi \alpha \alpha_s e_q^2}{\hat s^2} \frac{\hat u}{\hat s} .
\end{align}

\subsection{Dihadron helicity correlation in photon-nucleus collisions}

We assume the collinear factorization framework since we have integrated over the relative transverse momentum between the final state hadrons. Therefore, the cross-section of dihadron production is obtained by convoluting the dijet cross-section with two collinear fragmentation functions. 

We first consider the dijet production process in SIDIS which is given by $e + A \to e' + c + d + X$ with $c$ and $d$ denoting the flavor of two final state parton. We work in the photon-nucleus collinear frame with nucleus moving along the minus light-cone direction. The momenta of incoming and outgoing electrons are represented by $l_{1,2}$ respectively. The momentum of the intermediate spacelike photon is labeled as $q \equiv l_1-l_2$, which satisfies $q^2=-Q^2$. The per nucleon momentum is denoted as $P_n$. Thus, we can define Lorentz scalars usually used in the language of DIS physics: $x_{\rm Bj} = Q^2/2 P_n \cdot q$ and $y = P_n\cdot q/P_n\cdot l$. To simplify the discussion, we employ the factorized approach to compute the cross-section, which is given by 
\begin{align}
\frac{d\sigma^{e+A \to e' + c + d + X}_{\lambda_c \lambda_d}}{dx_{\rm Bj} dQ^2 dy_c d^2 \bm{k}_{Tc} dy_d d^2 \bm{k}_{Td}} = \sum_{\mathcal{S}=L,T} G_{\gamma^*,\mathcal{S}} (x_{\rm Bj}, Q^2) 
\sum_b x_b f_b (x_b) 
\frac{1}{\pi} 
\frac{d\hat \sigma^{\gamma^*_{\mathcal{S}} + b \to c + d}_{\lambda_c \lambda_d}}{d \hat t}
\delta^2 (\bm{k}_{Tc} + \bm{k}_{Td})
\delta(1-\frac{k_c^+ + k_d^+}{q^+}),
\end{align}
where $\mathcal{S}=L,T$ stands for the longitudinal and transverse virtual photons and $f_b(x_b)$ is the collinear parton distribution function with $x_b$ being the longitudinal momentum fraction (notice that $x_{\rm Bj}$ and $x_b$ are different variables). Here, $\lambda_{c,d}$ represents the helicity of final state partons, $y_{c,d}$ represents the rapidity and $\bm{k}_{Tc}=-\bm{k}_{Td}$ represents the transverse momentum in the collinear factorization. $G_{\gamma^*,\mathcal{S}}$ is the photon flux whose dimension is $-2$ with longitudinal and transverse sectors being given by
\begin{align}
&
G_{\gamma^*,T} (x_{\rm Bj}, Q^2) = \frac{\alpha}{2\pi Q^2 x_{\rm Bj}} [1 + (1-y)^2],
\\
&
G_{\gamma^*,L} (x_{\rm Bj}, Q^2) = \frac{\alpha}{\pi Q^2 x_{\rm Bj}} (1 - y).
\end{align}
Notice that our factorized approach is akin to that in Ref.~\cite{Caucal:2023nci} and we have taken our photon flux from Ref.~\cite{Caucal:2023nci}. However, Ref.~\cite{Caucal:2023nci} is interested in the gluon saturation physics and therefore employed the color glass condensate effective theory to compute the photon-nucleus interaction. Our paper is interested in the spin physics in the moderate-$x$ region and therefore adopts the collinear factorization.

As mentioned in the previous section, the interference between virtual photons with different helicities in the amplitude and conjugate amplitude disappears in the context that the azimuthal angle between leptonic plane and hadronic plane has been integrated over \cite{Jezuita-Dabrowska:2002tjg}. 

Convoluting with the helicity dependent fragmentation function ${\cal D}_{c \to h_1} (z_1, \lambda_c, \lambda_{1}) = D_{1,c} (z_1) + \lambda_c \lambda_{1} G_{1L,c} (z_1)$ and ${\cal D}_{d \to h_2} (z_2, \lambda_d, \lambda_{2}) = D_{1,d} (z_2) + \lambda_d \lambda_{2} G_{1L,d} (z_2)$ with $z_{1,2}$ the momentum fractions and $\lambda_{1,2}$ the helicities of final state hadrons, we obtain
\begin{align}
\frac{d\sigma^{\rm EIC}_{\lambda_{h_1} \lambda_{h_2}}}{dx_{\rm Bj} dQ^2 dy_{1} d^2 \bm{p}_{T1} dy_{2} d^2 \bm{p}_{T2}} 
&
=
\int \frac{dz_1}{z_1^2} \frac{dz_2}{z_2^2}
\sum_{\mathcal{S}=L,T} G_{\gamma^*,\mathcal{S}} (x_{\rm Bj}, Q^2) 
\sum_{b,c,d,\lambda_c,\lambda_d} x_b f_b (x_b) 
\frac{1}{\pi} 
\frac{d\hat \sigma^{\gamma^*_{\mathcal{S}} + b \to c + d}_{\lambda_c \lambda_d}}{d \hat t}
\nonumber \\
&\times 
{\cal D}_{c} (z_1, \lambda_c, \lambda_{h_1})
{\cal D}_{d} (z_2, \lambda_d, \lambda_{h_2})
\delta^2 \left(\frac{\bm{p}_{T1}}{z_1} + \frac{\bm{p}_{T2}}{z_2}\right)
\delta \left(\frac{|\bm{p}_{T1}|}{\sqrt{2}z_1 q^+} (e^{y_1} + e^{y_2}) - 1 \right)
\nonumber \\
&
 + \{ c \leftrightarrow d \}.
\label{eq:eic-cs}
\end{align}
Here, $\lambda_{1,2}$ represents the helicities of final state hadrons, $\bm{p}_{T1,2}$ represents the transverse momenta. Since we work in the collinear factorization, the hadron rapidity equals the parton rapidity, i.e., $y_1=y_c$ and $y_2=y_d$. Furthermore, we choose the Breit frame where the virtual photon has no energy. Thus the photon momentum is given by $q^\mu = (0, 0, 0, Q)$ in the Minkowski coordinates and $q^+ = - q^- = Q/\sqrt{2}$ in the light-cone coordinates. 

The factorization scale $\mu_f$ in parton distribution functions and fragmentation functions and the renormalization scale $\mu_r$ in the strong coupling $\alpha_s$ are not shown explicitly for simplicity. In the numerical evaluation, they should be considered. In general, the optimal choice is the typical hard scale of this process, which minimizes the higher order contribution. We can also estimate the theoretical uncertainty by varying the scales by a constant factor.

In the ultra-relativistic nucleus-nucleus collisions, the exchanged photons are quasireal. Therefore, the contribution from the longitudinal photon is eliminated. The differential cross-section thus reads
\begin{align}
\frac{d\sigma^{\rm UPC}_{\lambda_{h_1} \lambda_{h_2}}}{dy_{1} d^2 \bm{p}_{T1} dy_{2} d^2 \bm{p}_{T2}} 
&
=
\sum_{b,c,d,\lambda_c,\lambda_d} 
\int \frac{dz_1}{z_1^2} \frac{dz_2}{z_2^2}
x_{\gamma} f_{\gamma} (x_\gamma)
x_b f_b (x_b) \frac{1}{\pi} \frac{d\hat \sigma^{\gamma_T + b \to c + d}_{\lambda_c \lambda_d}}{d \hat t}
\nonumber \\
&\times 
{\cal D}_{c} (z_1, \lambda_c, \lambda_{h_1})
{\cal D}_{d} (z_2, \lambda_d, \lambda_{h_2})
\delta^2 (\frac{\bm{p}_{T1}}{z_1} + \frac{\bm{p}_{T2}}{z_2})  + \{ c \leftrightarrow d \}.
\end{align}
The coherent photon flux in the classic electrodynamics \cite{Jackson:1998nia} is given by
\begin{align}
x_\gamma f_\gamma (x_\gamma) = \frac{2Z^2 \alpha}{\pi} \left[ \zeta K_0 (\zeta) K_1(\zeta) - \frac{\zeta^2}{2} [K_1^2(\zeta) - K_0^2 (\zeta)] \right], \label{eq:photonflux}
\end{align}
where $Z$ is the atomic number of the large nucleus emitting quasireal photon and $\zeta = 2 x_\gamma M_p R_A$ with $x_\gamma$ the per nucleon momentum fraction carried by the photon, $M_p$ the proton mass and $R_A \sim 6$ fm the nucleus radius. Here $K_{0,1}$ are modified Bessel functions of the second kind.

\section{Numerical results}

Employing the DSV parametrization \cite{deFlorian:1997zj} for the polarized and unpolarized $\Lambda$ fragmentation functions, we can now make quantitative predictions for the future EIC experiment. We define the helicity correlation of the $\Lambda$-$\bar \Lambda$ pair as
\begin{align}
&
{\cal C}_{LL} (x_{\rm Bj}, Q^2)
= 
\frac{\int dy_1 dy_2 d^2 \bm{p}_{T1} d^2\bm{p}_{T2} \left[ \frac{d\sigma^{\rm EIC}_{++}}{d{P.S.}} + \frac{d\sigma^{\rm EIC}_{--}}{d{P.S.}} - \frac{d\sigma^{\rm EIC}_{+-}}{d{P.S.}} - \frac{d\sigma^{\rm EIC}_{-+}}{d{P.S.}} \right]}{\int dy_1 dy_2 d^2 \bm{p}_{T1} d^2\bm{p}_{T2} \left[ \frac{d\sigma^{\rm EIC}_{++}}{d{P.S.}} + \frac{d\sigma^{\rm EIC}_{--}}{d{P.S.}} + \frac{d\sigma^{\rm EIC}_{+-}}{d{P.S.}} + \frac{d\sigma^{\rm EIC}_{-+}}{d{P.S.}} \right]},
\end{align}
where $y_1$ and $\bm{p}_{T1}$ are the rapidity and transverse momentum of the $\Lambda$ hyperon and $y_2$ and $\bm{p}_{T2}$ are those of the $\bar \Lambda$ hyperon.

\begin{figure}[htb]
\includegraphics[width=0.31\textwidth]{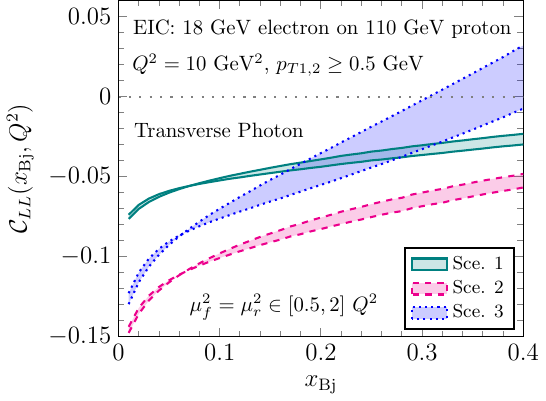}
\includegraphics[width=0.31\textwidth]{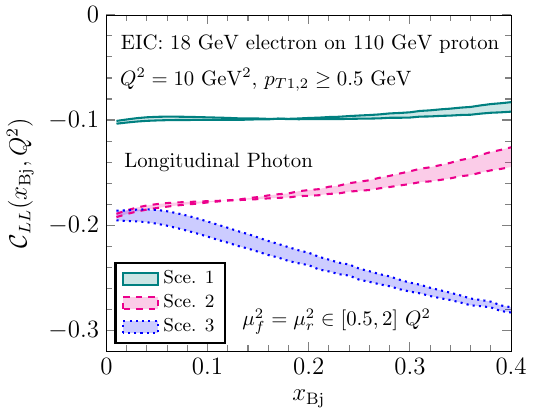}
\includegraphics[width=0.31\textwidth]{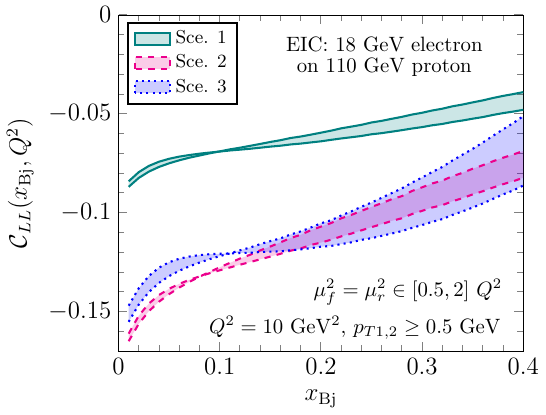}
\caption{Numerical predictions for the $\Lambda$-$\bar\Lambda$ helicity correlation as a function of $x_{\rm Bj}$ at $Q^2=10$ GeV$^2$ at the future EIC experiment. In the numerical evaluation, we have employed the DSV parametrization \cite{deFlorian:1997zj} for $\Lambda$ fragmentation functions and the CT14 parton distribution functions \cite{Dulat:2015mca}. Left: dihadron helicity correlation with a transverse photon; Middle: dihadron helicity correlation with a longitudinal photon; Right: dihadron helicity correlation for the future EIC experiment averaging over photon helicity.}
\label{fig:EIC}
\end{figure}

Taking the EIC kinematics, we present our numerical predictions for the helicity correlation of the $\Lambda$-$\bar\Lambda$ pair as a function of $x_{\rm Bj}$ at $Q^2=10$ GeV$^2$ in Fig.~\ref{fig:EIC}. The theoretical uncertainty is estimated by varying the factorization and renormalization scales by a few times around $Q^2$. To avoid nonperturbative contributions, we have required the transverse momenta of both final state hadrons to be larger than 0.5 GeV. The rapidities have been integrated in the whole kinematic region, i.e., $y_{1,2} \in [y_{\min}, y_{\max}]$. The lower limit $y_{\min}$ is set by requiring $x_b$ in Eq.~(\ref{eq:a2}) to be smaller than 1 and the upper limit $y_{\max}$ is determined by requiring $z_1$ in Eq.~(\ref{eq:a1}) to be smaller than 1.

We employ the DSV parameterization \cite{deFlorian:1997zj} to provide the unpolarized and polarized fragmentation function. For the unpolarized one, the DSV parameterization only provides the sum of $u\to \Lambda$ and $\bar u\to \Lambda$ fragmentation functions denoted by $D_{1}^{u/\bar u \to \Lambda} (z)$. To separate these two contributions, we resort to a phenomenological prescription, e.g., Refs.~\cite{DAlesio:2020wjq, Callos:2020qtu, Chen:2021hdn}. Our prescription is laid out as 
\begin{align}
& D_{1}^{u\to \Lambda} (z) = \frac{1+z}{2} D_{1}^{u/\bar u \to \Lambda} (z),
\\
& D_{1}^{\bar u\to \Lambda} (z) = \frac{1-z}{2} D_{1}^{u/\bar u \to \Lambda} (z).
\end{align}
For the polarized fragmentation function, Ref.~\cite{deFlorian:1997zj} does provide the separation of $u$ and $\bar u$. Additionally, it also provides three scenarios. The first scenario establishes on the naive quark model and therefore assumes that only the $s$ quark contributes to the $\Lambda$ polarization at the initial scale. The second scenario, on the other hand, assumes that the polarized fragmentation function of $u/d$ is negative. The third scenario is an ``extreme'' one assuming that $u$, $d$, and $s$ contribute equally. Notice that all three scenarios have assumed the isospin symmetry in the parametrization of $G_{1L}$. The isospin symmetry of Lambda fragmentation functions has been discussed in details in Ref.~\cite{Chen:2021hdn}, while Ref.~\cite{Chen:2021zrr} also demonstrated that EIC experiment has a great potential in testing this symmetry. Their conclusion has also been confirmed by Refs.~\cite{DAlesio:2022brl, DAlesio:2023ozw}. In this work, we skip the discussion of isospin symmetry and make predictions with three isospin symmetric scenarios. We show the potential of dihadron helicity correlation in distinguishing which scenario describes hadronization process the best.

Although it is not possible to distinguish different photon polarizations, we still present the correlations in different processes separately, since they exhibit distinct features. For the transverse photon, the $\gamma_T^* g$ channel contributes to the negative correlation, while the $\gamma_T^* q$ channel contributes to the positive correlation. At small $x_{\rm Bj}$, there are more contributions from the gluon channel (recalling $x_{\rm Bj}$ is also the lower limit of the parton momentum fraction $x_{b}$). With increasing $x_{\rm Bj}$, the $\gamma_T^* q$ channel becomes more and more significant. Therefore, due to the competition between these two channels, the helicity correlation becomes smaller in absolute magnitude with increasing $x_{\rm Bj}$. However, the spin transfer of $u$ and $d$ quarks is assumed to be very small in Scenarios 1 and 2. Therefore, although the $\gamma_T^* q$ with $q=u, d$ channel dominates the unpolarized cross-section at large $x_{\rm Bj}$, its contribution to the dihadron helicity correlation is still not on par with that of the $\gamma_T^* g$ channel. On the other hand, Scenario 3 assumes that $G_{1L,q}$ remains the same for $u$, $d$, and $s$ quarks. Therefore, the $\gamma_T^* q$ channel also has a significant contribution to the dihadron helicity correlation. As shown in Fig.~\ref{fig:EIC}, in Scenario 3, the $\gamma_T^* q$ channel even overcomes the $\gamma_T^* g$ channel and the correlation becomes positive at $x_{\rm Bj} > 0.4$. However, for the longitudinal photon, both $\gamma_L^* g$ and $\gamma_L^* q$ channels amount to negative correlations. Therefore, the correlation is negative definite and the magnitude is much larger than that for the transverse photon. 

\begin{figure}[htb]
\includegraphics[width=0.31\textwidth]{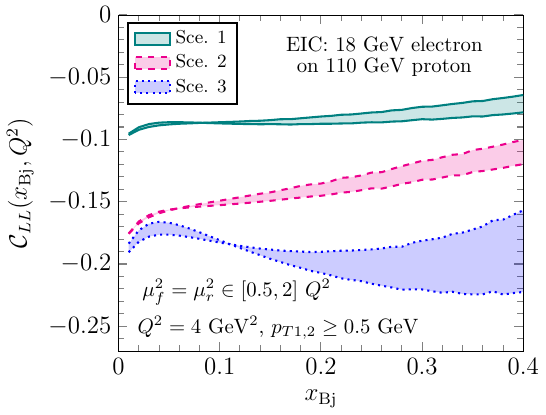}
\includegraphics[width=0.31\textwidth]{CLL_SUM}
\includegraphics[width=0.31\textwidth]{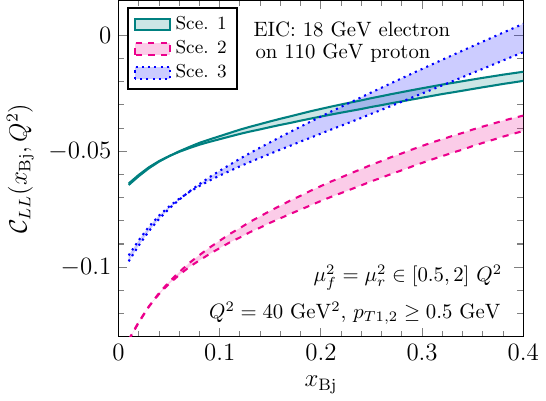}
\caption{Dihadron helicity correlation as a function of $x_{\rm Bj}$ at different $Q^2$. Here we only show numeric predictions averaged over photon polarization.}
\label{fig:EIC-2}
\end{figure}

Furthermore, the longitudinal photon contribution becomes significant at high $Q^2$ and disappears at $Q^2 \to 0$. Therefore, by measuring the dihadron helicity correlation at different virtuality, we can adjust relative contributions from transverse and longitudinal sectors. Notice that the hard scale of the partonic process becomes the partonic center of mass energy in the $Q^2 \to 0$ limit. Thus, we shall require the transverse momentum of the final state hadron to be a bit larger to avoid the contamination of nonperturbative physics. The relative contribution from the $\gamma^* q \to qg$ channel is mainly controlled by $x_{\rm Bj}$, which grants the SIDIS process more resolution power on gluon fragmentation functions compared with the electron positron annihilation process. On top of that, the relative contribution from the transverse and longitudinal photons can be tuned by varying $Q^2$. The sign change of the partonic helicity correlation of $\gamma^*_L q$ and $\gamma^*_T q$ channels will result in a nontrivial $Q^2$ dependence of dihadron helicity correlation. This feature thus can be utilized as a tool to understand the flavor dependence of the spin transfer $G_{1L}$. Particularly, it allows us to probe the fragmentation function of circularly polarized gluons. As shown in Fig.~\ref{fig:EIC-2}, three scenarios of the DSV parametrization predict different dihadron helicity correlations at different $x_{\rm Bj}$ and different $Q^2$.

\begin{figure}[htb]
\includegraphics[width=0.32\textwidth]{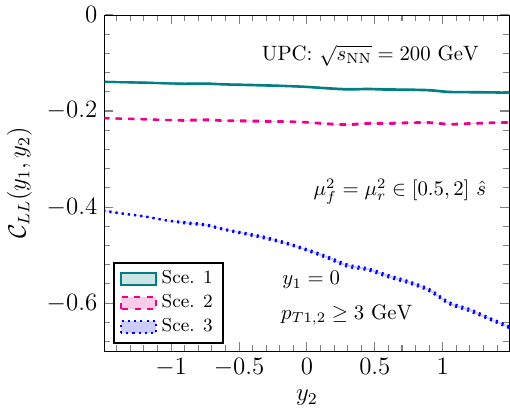}
\includegraphics[width=0.31\textwidth]{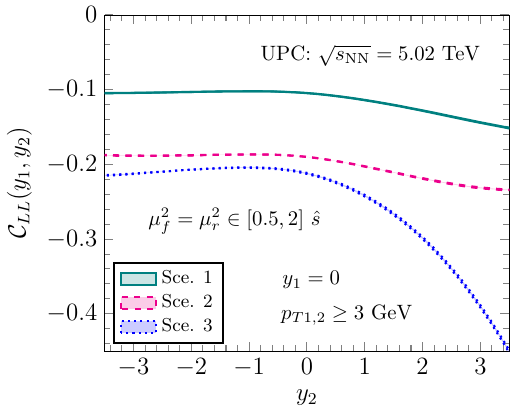}
\caption{Dihadron helicity correlation as a function of $\bar\Lambda$ rapidity $y_2$ in UPC. We have chosen the photon-going direction as the positive rapidity direction. EPPS nuclear parton distribution function \cite{Eskola:2021nhw} for the struck nucleus has been employed in the numerical evaluation. Left: predictions for the RHIC AuAu UPC experiment. Right: predictions for the LHC PbPb UPC experiment.}
\label{fig:UPC}
\end{figure}

For the UPC process discussed in this paper, we also define the $\Lambda$-$\bar\Lambda$ helicity correlation ${\cal C}_{LL}$ as
\begin{align}
&
{\cal C}_{LL} (y_1, y_2)
= 
\frac{\int d^2 \bm{p}_{T1} d^2\bm{p}_{T2} \left[ \frac{d\sigma^{\rm UPC}_{++}}{d{P.S.}} + \frac{d\sigma^{\rm UPC}_{--}}{d{P.S.}} - \frac{d\sigma^{\rm UPC}_{+-}}{d{P.S.}} - \frac{d\sigma^{\rm UPC}_{-+}}{d{P.S.}} \right]}{\int d^2 \bm{p}_{T1} d^2\bm{p}_{T2} \left[ \frac{d\sigma^{\rm UPC}_{++}}{d{P.S.}} + \frac{d\sigma^{\rm UPC}_{--}}{d{P.S.}} + \frac{d\sigma^{\rm UPC}_{+-}}{d{P.S.}} + \frac{d\sigma^{\rm UPC}_{-+}}{d{P.S.}} \right]}.
\end{align}
We show the numerical results as a function of $y_2$ for $y_1=0$ in Fig.~\ref{fig:UPC}. Since only one of the large nuclei remains in one piece, we are empowered to figure out the photon-going direction, which is defined as the positive/forward rapidity direction in this work. The forward and backward directions are thus asymmetric. 

Since the coherent photons are quasireal, the only hard scale in this process is the partonic center of mass energy $\sqrt{\hat s}$. Therefore, we require $p_{T1,2}\ge 3$ GeV to ensure that we are working in the perturbative regime. In a more forward rapidity, the kinematics forces $x_{\gamma} z_{1,2}$ to be larger. However, the coherent photon flux drops exponentially at large $x_{\gamma}$. Therefore, this kinematic requirement effectively leads to larger typical $z_{1,2}$. Measuring the dihadron helicity correlation in different rapidities can be utilized to probe the $z$ dependence of the longitudinal spin transfer.

We have varied the factorization and renormalization scales by a few times of the partonic center of mass energy $\sqrt{\hat s}$. It is worth noting that the scale dependence is very small compared with that for the EIC experiment. The reason is twofold. First, according to the kinematics specified in this paper, the typical factorization scale in the UPC experiment is usually much larger than that in the EIC experiment. The DGLAP evolution effect becomes milder at larger scale. Second, in the EIC experiment, $\gamma^* g$ and $\gamma^* q$ are equally important, and the relative contribution from each channel is scale dependent. On the other hand, in the UPC experiment, the dominant contribution comes from the $\gamma g$ channel. Therefore, there is a new source of scale dependence in the EIC experiment.

\section{Summary}

In this paper, we investigate the helicity correlation of two almost back-to-back dihadron produced in photon nucleus collisions. We first derive the helicity amplitudes for the photon-parton scattering with a virtual photon. Taking the $Q^2 = 0$ limit, our helicity amplitudes coincide with those derived in Ref.~\cite{Gastmans:1990xh} for real photons. Furthermore, convoluting the helicity-dependent partonic cross section combined with the helicity-dependent fragmentation functions of $\Lambda$ hyperons, we make quantitative predictions for the future EIC experiment and the current UPC experiment. 

Numerical results show that the dihadron helicity correlation is at around $10\% \sim 30\%$ varying with kinematics. In light of the sizable signal, it is quite plausible that the dihadron helicity correlation could be eventually measured in experiments. The experimental measurements will reveal more information on the flavor dependence of the longitudinal spin transfer $G_{1L}$ and shed more light on the fragmentation function of circularly polarized gluons.

\section*{Acknowledgments}
We thank T. Liu and Y.K. Song for helpful discussions. S.Y. Wei is supported by the Shandong Province Natural Science Foundation under Grant No.~2023HWYQ-011 and the Taishan fellowship of Shandong Province for junior scientists. H. Dong is also supported in part by the National Natural Science Foundation of China (approval no. 12175118).

\appendix

\section{Useful relations}

In this appendix, we supplement with additional details regarding the numerical calculation for the EIC experiment. We work in the Breit frame and present the relations among different variables. We compute the dihadron helicity correlation with given $x_{\rm Bj}$, $Q^2$, and $s=2 P_n \cdot l_1$ and integrate over rapidities and transverse momenta of final state hadrons. According to the momentum conservation, we find that $q^+$ and $y$ are solely determined by the kinematics and obtain
\begin{align}
&
q^+ = \frac{Q}{\sqrt{2}} = \frac{|\bm{p}_{T1}|}{\sqrt{2}z_1 } (e^{y_1} + e^{y_2}), \label{eq:a1}
\\
&
y = \frac{P_n \cdot q}{P_n \cdot l_1} = \frac{1}{x_{\rm Bj}} \frac{Q^2}{s}.
\end{align}
While the first delta function in Eq.~(\ref{eq:eic-cs}) cancels the integral over $\bm{p}_{T2}$, the second one terminates the $z_1$ integral and sets $z_1 = |\bm{p}_{T1}|(e^{y_1} + e^{y_2})/Q$.

Furthermore, the momentum fraction $x_b$ can further be evaluated from the conservation of light-cone minus momentum. From the following relation, 
\begin{align}
& 
Q^2 = - 2 q^+ q^-,
&&
x_{\rm Bj} = \frac{Q^2}{2P_n \cdot q} = - \frac{q^-}{P_n^-},
\end{align}
it is straightforward to obtain
\begin{align}
x_b = \frac{1}{P_n^-} \left[ \frac{|\bm{p}_{T1}|}{\sqrt{2}z_1} (e^{-y_1} + e^{-y_2}) - q^- \right]
= \frac{1}{P_n^-} \left[ \frac{|\bm{p}_{T1}|}{\sqrt{2}z_1} (e^{-y_1} + e^{-y_2}) \right] + x_{\rm Bj}.\label{eq:a2}
\end{align}

\section{An estimate of the mass effect}

\begin{figure}[h!]\centering
\includegraphics[width=0.6\textwidth]{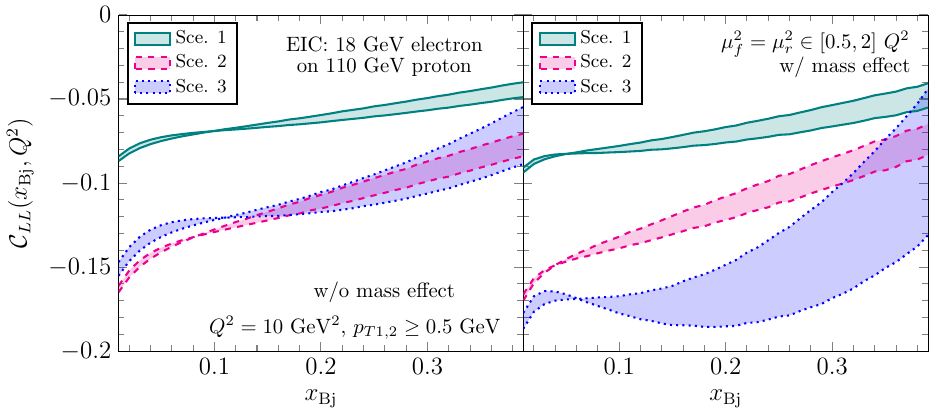}
\includegraphics[width=0.6\textwidth]{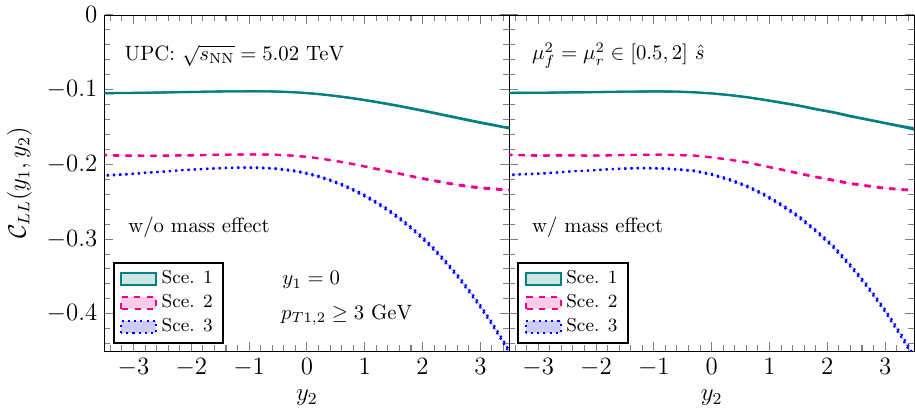}
\caption{A numerical estimate of the mass effect in EIC (upper) and UPC (lower) experiments.}
\label{fig:mass}
\end{figure}

In this paper, we have adopted the leading twist approximation in the collinear factorization. Therefore, corrections from higher twist parton distribution functions and fragmentation functions are neglected. This approximation is appropriate since $M_{\Lambda}^2/\mu^2 \ll 1$ with $M_{\Lambda}$ the Lambda mass and $\mu$ the typical hard scale. 

Beyond those genuine higher twist corrections, there are also kinematic higher twist effects due to the mass of $\Lambda$. These kinematic higher twist effects should also be small as long as $p_T \gg M_\Lambda$. While this is true in the UPC experiment, the phase space for high $p_T$ hadron production is very limited in the EIC experiment. A systemic and consistent consideration of the mass effect is complicated. In this section, we estimate the impact of the mass effect with a simple prescription. 

For the EIC experiments: First, we replace the second delta function in Eq.~(\ref{eq:eic-cs}) by
\begin{align}
&
\delta \left(\frac{|\bm{p}_{T1}|}{\sqrt{2}z_1 q^+} (e^{y_1} + e^{y_2}) - 1 \right)
&&
\Rightarrow
&&
\delta \left(\frac{m_{T1}}{\sqrt{2}z_1 q^+} (e^{y_1} + e^{y_2}) - 1 \right),
\end{align}
where $m_{T1} \equiv \sqrt{\bm{p}_{T1}^2 + M_{\Lambda}^2}$ is the transverse mass. Second, we replace all the $|\bm{p}_{T1}|$'s in Appendix A with $m_{T1}$. A similar modification is also implemented in the calculation for the UPC experiment.

The numerical results based on the above prescription are presented in Fig.~\ref{fig:mass}. As expected, the mass effect in the UPC experiments is negligible. On the other hand, it has a visible impact in the EIC experiments where the typical transverse momentum is around $M_{\Lambda}$.


\begin{thebibliography}{9}
\bibitem{Metz:2016swz}
A.~Metz and A.~Vossen,
Prog. Part. Nucl. Phys. \textbf{91} (2016), 136-202
doi:10.1016/j.ppnp.2016.08.003
[arXiv:1607.02521 [hep-ex]].

\bibitem{Chen:2023kqw}
K.~B.~Chen, T.~Liu, Y.~K.~Song and S.~Y.~Wei,
Particles \textbf{6} (2023) no.2, 515-545
doi:10.3390/particles6020029
[arXiv:2307.02874 [hep-ph]].

\bibitem{Boussarie:2023izj}
R.~Boussarie, M.~Burkardt, M.~Constantinou, W.~Detmold, M.~Ebert, M.~Engelhardt, S.~Fleming, L.~Gamberg, X.~Ji and Z.~B.~Kang, \textit{et al.}
[arXiv:2304.03302 [hep-ph]].

\bibitem{Binnewies:1994ju}
J.~Binnewies, B.~A.~Kniehl and G.~Kramer,
Z. Phys. C \textbf{65} (1995), 471-480
doi:10.1007/BF01556135
[arXiv:hep-ph/9407347 [hep-ph]].

\bibitem{Kniehl:2000fe}
B.~A.~Kniehl, G.~Kramer and B.~Potter,
Nucl. Phys. B \textbf{582} (2000), 514-536
doi:10.1016/S0550-3213(00)00303-5
[arXiv:hep-ph/0010289 [hep-ph]].

\bibitem{Kneesch:2007ey}
T.~Kneesch, B.~A.~Kniehl, G.~Kramer and I.~Schienbein,
Nucl. Phys. B \textbf{799} (2008), 34-59
doi:10.1016/j.nuclphysb.2008.02.015
[arXiv:0712.0481 [hep-ph]].

\bibitem{Kretzer:2000yf}
S.~Kretzer,
Phys. Rev. D \textbf{62} (2000), 054001
doi:10.1103/PhysRevD.62.054001
[arXiv:hep-ph/0003177 [hep-ph]].

\bibitem{Albino:2005me}
S.~Albino, B.~A.~Kniehl and G.~Kramer,
Nucl. Phys. B \textbf{725} (2005), 181-206
doi:10.1016/j.nuclphysb.2005.07.010
[arXiv:hep-ph/0502188 [hep-ph]].

\bibitem{Albino:2005mv}
S.~Albino, B.~A.~Kniehl and G.~Kramer,
Nucl. Phys. B \textbf{734} (2006), 50-61
doi:10.1016/j.nuclphysb.2005.11.006
[arXiv:hep-ph/0510173 [hep-ph]].

\bibitem{Albino:2008fy}
S.~Albino, B.~A.~Kniehl and G.~Kramer,
Nucl. Phys. B \textbf{803} (2008), 42-104
doi:10.1016/j.nuclphysb.2008.05.017
[arXiv:0803.2768 [hep-ph]].

\bibitem{deFlorian:2007aj}
D.~de Florian, R.~Sassot and M.~Stratmann,
Phys. Rev. D \textbf{75} (2007), 114010
doi:10.1103/PhysRevD.75.114010
[arXiv:hep-ph/0703242 [hep-ph]].

\bibitem{deFlorian:2007ekg}
D.~de Florian, R.~Sassot and M.~Stratmann,
Phys. Rev. D \textbf{76} (2007), 074033
doi:10.1103/PhysRevD.76.074033
[arXiv:0707.1506 [hep-ph]].

\bibitem{Hirai:2007cx}
M.~Hirai, S.~Kumano, T.~H.~Nagai and K.~Sudoh,
Phys. Rev. D \textbf{75} (2007), 094009
doi:10.1103/PhysRevD.75.094009
[arXiv:hep-ph/0702250 [hep-ph]].

\bibitem{Aidala:2010bn}
C.~A.~Aidala, F.~Ellinghaus, R.~Sassot, J.~P.~Seele and M.~Stratmann,
Phys. Rev. D \textbf{83} (2011), 034002
doi:10.1103/PhysRevD.83.034002
[arXiv:1009.6145 [hep-ph]].

\bibitem{deFlorian:2014xna}
D.~de Florian, R.~Sassot, M.~Epele, R.~J.~Hern\'andez-Pinto and M.~Stratmann,
Phys. Rev. D \textbf{91} (2015) no.1, 014035
doi:10.1103/PhysRevD.91.014035
[arXiv:1410.6027 [hep-ph]].

\bibitem{deFlorian:2017lwf}
D.~de Florian, M.~Epele, R.~J.~Hernandez-Pinto, R.~Sassot and M.~Stratmann,
Phys. Rev. D \textbf{95} (2017) no.9, 094019
doi:10.1103/PhysRevD.95.094019
[arXiv:1702.06353 [hep-ph]].

\bibitem{Bertone:2017tyb}
V.~Bertone \textit{et al.} [NNPDF],
Eur. Phys. J. C \textbf{77} (2017) no.8, 516
doi:10.1140/epjc/s10052-017-5088-y
[arXiv:1706.07049 [hep-ph]].

\bibitem{Khalek:2021gxf}
R.~A.~Khalek \textit{et al.} [MAP (Multi-dimensional Analyses of Partonic distributions)],
Phys. Rev. D \textbf{104} (2021) no.3, 034007
doi:10.1103/PhysRevD.104.034007
[arXiv:2105.08725 [hep-ph]].

\bibitem{arXiv:1905.03788}
N.~Sato \textit{et al.} [JAM],
Phys. Rev. D \textbf{101} (2020) no.7, 074020
doi:10.1103/PhysRevD.101.074020
[arXiv:1905.03788 [hep-ph]].

\bibitem{arXiv:2101.04664}
E.~Moffat \textit{et al.} [Jefferson Lab Angular Momentum (JAM)],
Phys. Rev. D \textbf{104} (2021) no.1, 016015
doi:10.1103/PhysRevD.104.016015
[arXiv:2101.04664 [hep-ph]].

\bibitem{arXiv:2210.06078}
M.~Czakon, T.~Generet, A.~Mitov and R.~Poncelet,
JHEP \textbf{03} (2023), 251
doi:10.1007/JHEP03(2023)251
[arXiv:2210.06078 [hep-ph]].

\bibitem{Augustin:1978wf}
J.~E.~Augustin and F.~M.~Renard,
Nucl. Phys. B \textbf{162} (1980), 341
doi:10.1016/0550-3213(80)90269-2

\bibitem{Gustafson:1992iq}
G.~Gustafson and J.~Hakkinen,
Phys. Lett. B \textbf{303} (1993), 350-354
doi:10.1016/0370-2693(93)91444-R

\bibitem{ALEPH:1996oew}
D.~Buskulic \textit{et al.} [ALEPH],
Phys. Lett. B \textbf{374} (1996), 319-330
doi:10.1016/0370-2693(96)00300-0

\bibitem{OPAL:1997oem}
K.~Ackerstaff \textit{et al.} [OPAL],
Eur. Phys. J. C \textbf{2} (1998), 49-59
doi:10.1007/s100520050123
[arXiv:hep-ex/9708027 [hep-ex]].

\bibitem{E665:1999fso}
M.~R.~Adams \textit{et al.} [E665],
Eur. Phys. J. C \textbf{17} (2000), 263-267
doi:10.1007/s100520000493
[arXiv:hep-ex/9911004 [hep-ex]].

\bibitem{HERMES:1999buc}
A.~Airapetian \textit{et al.} [HERMES],
Phys. Rev. D \textbf{64} (2001), 112005
doi:10.1103/PhysRevD.64.112005
[arXiv:hep-ex/9911017 [hep-ex]].

\bibitem{NOMAD:2000wdf}
P.~Astier \textit{et al.} [NOMAD],
Nucl. Phys. B \textbf{588} (2000), 3-36
doi:10.1016/S0550-3213(00)00503-4

\bibitem{NOMAD:2001iup}
P.~Astier \textit{et al.} [NOMAD],
Nucl. Phys. B \textbf{605} (2001), 3-14
doi:10.1016/S0550-3213(01)00181-X
[arXiv:hep-ex/0103047 [hep-ex]].

\bibitem{HERMES:2006lro}
A.~Airapetian \textit{et al.} [HERMES],
Phys. Rev. D \textbf{74} (2006), 072004
doi:10.1103/PhysRevD.74.072004
[arXiv:hep-ex/0607004 [hep-ex]].

\bibitem{COMPASS:2009nhs}
M.~Alekseev \textit{et al.} [COMPASS],
Eur. Phys. J. C \textbf{64} (2009), 171-179
doi:10.1140/epjc/s10052-009-1143-7
[arXiv:0907.0388 [hep-ex]].

\bibitem{deFlorian:1997zj}
D.~de Florian, M.~Stratmann and W.~Vogelsang,
Phys. Rev. D \textbf{57} (1998), 5811-5824
doi:10.1103/PhysRevD.57.5811
[arXiv:hep-ph/9711387 [hep-ph]].

\bibitem{Chen:1994ar}
K.~Chen, G.~R.~Goldstein, R.~L.~Jaffe and X.~D.~Ji,
Nucl. Phys. B \textbf{445} (1995), 380-398
doi:10.1016/0550-3213(95)00193-V
[arXiv:hep-ph/9410337 [hep-ph]].

\bibitem{Zhang:2023ugf}
H.~C.~Zhang and S.~Y.~Wei,
Phys. Lett. B \textbf{839} (2023), 137821
doi:10.1016/j.physletb.2023.137821
[arXiv:2301.04096 [hep-ph]].

\bibitem{Li:2023qgj}
X.~Li, Z.~X.~Chen, S.~Cao and S.~Y.~Wei,
Phys. Rev. D \textbf{109} (2024) no.1, 014035
doi:10.1103/PhysRevD.109.014035
[arXiv:2309.09487 [hep-ph]].

\bibitem{Gong:2021bcp}
W.~Gong, G.~Parida, Z.~Tu and R.~Venugopalan,
Phys. Rev. D \textbf{106} (2022) no.3, L031501
doi:10.1103/PhysRevD.106.L031501
[arXiv:2107.13007 [hep-ph]].

\bibitem{Vanek:2023oeo}
J.~Vanek [STAR],
[arXiv:2307.07373 [nucl-ex]].

\bibitem{Tu:2023few}
Z.~Tu,
[arXiv:2308.09127 [hep-ph]].

\bibitem{DeCausmaecker:1981wzb}
P.~De Causmaecker, R.~Gastmans, W.~Troost and T.~T.~Wu,
Phys. Lett. B \textbf{105} (1981), 215
doi:10.1016/0370-2693(81)91025-X

\bibitem{Gastmans:1990xh}
R.~Gastmans and T.~T.~Wu,
Int. Ser. Monogr. Phys. \textbf{80} (1990), 1-648

\bibitem{Anselmino:2005sh}
M.~Anselmino, M.~Boglione, U.~D'Alesio, E.~Leader, S.~Melis and F.~Murgia,
Phys. Rev. D \textbf{73} (2006), 014020
doi:10.1103/PhysRevD.73.014020
[arXiv:hep-ph/0509035 [hep-ph]].

\bibitem{Anselmino:2011ch}
M.~Anselmino, M.~Boglione, U.~D'Alesio, S.~Melis, F.~Murgia, E.~R.~Nocera and A.~Prokudin,
Phys. Rev. D \textbf{83} (2011), 114019
doi:10.1103/PhysRevD.83.114019
[arXiv:1101.1011 [hep-ph]].

\bibitem{DAlesio:2021dcx}
U.~D'Alesio, F.~Murgia and M.~Zaccheddu,
JHEP \textbf{10} (2021), 078
doi:10.1007/JHEP10(2021)078
[arXiv:2108.05632 [hep-ph]].

\bibitem{AbdulKhalek:2021gbh}
R.~Abdul Khalek, A.~Accardi, J.~Adam, D.~Adamiak, W.~Akers, M.~Albaladejo, A.~Al-bataineh, M.~G.~Alexeev, F.~Ameli and P.~Antonioli, \textit{et al.}
Nucl. Phys. A \textbf{1026} (2022), 122447
doi:10.1016/j.nuclphysa.2022.122447
[arXiv:2103.05419 [physics.ins-det]].

\bibitem{Jaffe:1996wp}
R.~L.~Jaffe,
Phys. Rev. D \textbf{54} (1996) no.11, R6581-R6585
doi:10.1103/PhysRevD.54.R6581
[arXiv:hep-ph/9605456 [hep-ph]].

\bibitem{Kotzinian:1997vd}
A.~Kotzinian, A.~Bravar and D.~von Harrach,
Eur. Phys. J. C \textbf{2} (1998), 329-337
doi:10.1007/s100520050142
[arXiv:hep-ph/9701384 [hep-ph]].

\bibitem{deFlorian:1997kt}
D.~de Florian, M.~Stratmann and W.~Vogelsang,
[arXiv:hep-ph/9710410 [hep-ph]].

\bibitem{Ashery:1999am}
D.~Ashery and H.~J.~Lipkin,
Phys. Lett. B \textbf{469} (1999), 263-269
doi:10.1016/S0370-2693(99)01229-0
[arXiv:hep-ph/9908355 [hep-ph]].

\bibitem{deFlorian:1998ba}
D.~de Florian, M.~Stratmann and W.~Vogelsang,
Phys. Rev. Lett. \textbf{81} (1998), 530-533
doi:10.1103/PhysRevLett.81.530
[arXiv:hep-ph/9802432 [hep-ph]].

\bibitem{Jager:2002xm}
B.~Jager, A.~Schafer, M.~Stratmann and W.~Vogelsang,
Phys. Rev. D \textbf{67} (2003), 054005
doi:10.1103/PhysRevD.67.054005
[arXiv:hep-ph/0211007 [hep-ph]].

\bibitem{Xu:2002hz}
Q.~h.~Xu, C.~x.~Liu and Z.~t.~Liang,
Phys. Rev. D \textbf{65} (2002), 114008
doi:10.1103/PhysRevD.65.114008
[arXiv:hep-ph/0204318 [hep-ph]].

\bibitem{Xu:2005ru}
Q.~h.~Xu, Z.~t.~Liang and E.~Sichtermann,
Phys. Rev. D \textbf{73} (2006), 077503
doi:10.1103/PhysRevD.73.077503
[arXiv:hep-ph/0511061 [hep-ph]].

\bibitem{Jezuita-Dabrowska:2002tjg}
U.~Jezuita-Dabrowska and M.~Krawczyk,
[arXiv:hep-ph/0211112 [hep-ph]].

\bibitem{Field:1989uq}
R.~D.~Field,
Front. Phys. \textbf{77} (1989), 1-366.

\bibitem{Huang:2003uy}
H.~W.~Huang and T.~Morii,
Phys. Rev. D \textbf{68} (2003), 014016
doi:10.1103/PhysRevD.68.014016
[arXiv:hep-ph/0305132 [hep-ph]].

\bibitem{Caucal:2023nci}
P.~Caucal, F.~Salazar, B.~Schenke, T.~Stebel and R.~Venugopalan,
JHEP \textbf{08} (2023), 062
doi:10.1007/JHEP08(2023)062
[arXiv:2304.03304 [hep-ph]].


\bibitem{Jackson:1998nia}
J.~D.~Jackson,
Wiley, 1998,
ISBN 978-0-471-30932-1

\bibitem{Dulat:2015mca}
S.~Dulat, T.~J.~Hou, J.~Gao, M.~Guzzi, J.~Huston, P.~Nadolsky, J.~Pumplin, C.~Schmidt, D.~Stump and C.~P.~Yuan,
Phys. Rev. D \textbf{93} (2016) no.3, 033006
doi:10.1103/PhysRevD.93.033006
[arXiv:1506.07443 [hep-ph]].


\bibitem{DAlesio:2020wjq}
U.~D'Alesio, F.~Murgia and M.~Zaccheddu,
Phys. Rev. D \textbf{102} (2020) no.5, 054001
doi:10.1103/PhysRevD.102.054001
[arXiv:2003.01128 [hep-ph]].


\bibitem{Callos:2020qtu}
D.~Callos, Z.~B.~Kang and J.~Terry,
Phys. Rev. D \textbf{102} (2020) no.9, 096007
doi:10.1103/PhysRevD.102.096007
[arXiv:2003.04828 [hep-ph]].

\bibitem{Chen:2021hdn}
K.~B.~Chen, Z.~T.~Liang, Y.~L.~Pan, Y.~K.~Song and S.~Y.~Wei,
Phys. Lett. B \textbf{816} (2021), 136217
doi:10.1016/j.physletb.2021.136217
[arXiv:2102.00658 [hep-ph]].

\bibitem{Chen:2021zrr}
K.~b.~Chen, Z.~t.~Liang, Y.~k.~Song and S.~y.~Wei,
Phys. Rev. D \textbf{105} (2022) no.3, 034027
doi:10.1103/PhysRevD.105.034027
[arXiv:2108.07740 [hep-ph]].

\bibitem{DAlesio:2022brl}
U.~D'Alesio, L.~Gamberg, F.~Murgia and M.~Zaccheddu,
JHEP \textbf{12} (2022), 074
doi:10.1007/JHEP12(2022)074
[arXiv:2209.11670 [hep-ph]].

\bibitem{DAlesio:2023ozw}
U.~D'Alesio, L.~Gamberg, F.~Murgia and M.~Zaccheddu,
Phys. Rev. D \textbf{108} (2023) no.9, 094004
doi:10.1103/PhysRevD.108.094004
[arXiv:2307.02359 [hep-ph]].

\bibitem{Eskola:2021nhw}
K.~J.~Eskola, P.~Paakkinen, H.~Paukkunen and C.~A.~Salgado,
Eur. Phys. J. C \textbf{82} (2022) no.5, 413
doi:10.1140/epjc/s10052-022-10359-0
[arXiv:2112.12462 [hep-ph]].
\end{thebibliography}
\end{document}